\documentclass[onecolumn,prl,showpacs,tightenlines]{revtex4}
\usepackage{epsfig,graphicx,times}
\usepackage{amstext}
\usepackage{amsmath}
\usepackage{amssymb}
\usepackage{graphicx}
\usepackage{latexsym}
\usepackage{bm}

\begin{document}

\title{ Finite-Temperature Scaling of Magnetic Susceptibility and Geometric Phase in the XY Spin Chain}
\author{H. T. Quan}
\affiliation{Theoretical Division, MS B213, Los Alamos National Laboratory, Los Alamos,
NM, 87545, U.S.A.}

\begin{abstract}
We study the magnetic susceptibility of 1D quantum XY model, and show that when the temperature approaches zero, the magnetic susceptibility exhibits the finite-temperature scaling
behavior. This scaling behavior of the magnetic susceptibility in 1D quantum XY model, due to the quantum-classical mapping, can be easily experimentally tested.
Furthermore, the universality in the critical properties of the magnetic susceptibility in quantum XY model is verified. Our study also reveals the close relation between the magnetic susceptibility and the geometric phase in some spin systems, 
where the quantum phase transitions are driven by an external magnetic field.
\end{abstract}

\maketitle

\emph{Introduction:} 
Quantum Phase Transitions (QPTs), which occur at absolute zero and are driven by zero-point quantum fluctuations,
are one of the most fascinating aspects of many-body systems. QPTs and related quantum critical phenomena have been a topic of tremendous interest in condensed matter physics and have been extensively studied in the past decade \cite {sachdev}. 
In recent work, quantum criticality has been characterized by using the methods and notions borrowed from quantum information science, such as the concurrence \cite {concurrence}, 
the entanglement entropy \cite {entanglemententropy}, geometric phase (GP) \cite{pachos}, Loschmidt Echo \cite{LE}, and quantum fidelity \cite{fidelity} in the place of traditional criteria,
 such as specific heat or magnetic susceptibility (MS). Most of these studies focus on the zero-temperature properties of the critical systems. 
In recent years, the finite-temperature properties of QPTs \cite{nature05,nonzero}, such as, thermal entanglement \cite{thermal} have begun to attract more attention. 
This is because, firstly, all experiments are confined to finite temperature. Thus, to experimentally verify 
the theoretical results, knowing only the zero-temperature properties of the quantum system
is not sufficient. Secondly, though genuine QPTs occur only at absolute zero, quantum criticality has profound influence on system properties up to a surprisingly high temperature \cite {nature05}. 
Interesting phenomena at finite temperature related to QPTs have been experimentally observed
in various systems, such as the heavy fermion system and the BEC \cite {review}.

On the other hand, it has been shown that a QPT in $d$ space dimensions is related to a classical transition in $d+z$ space dimensions \cite{sachdev, map}, where $z$ is the dynamical critical exponent. Under this quantum-classical mapping, the temperature
$T$ of quantum system maps onto an imaginary time direction: $\tau=-it/\hbar \in (0, 1/k_BT)$, where $\tau$ and $t$ are imaginary and real time \cite{map}. Accordingly accessing
 the QPT by reducing the temperature amounts to increasing the size of imaginary time dimension towards 
infinity, and leads to a divergence of the spatial correlation length $\xi$. This one-to-one mapping motivates us to study the finite-temperature properties of QPTs through its higher dimensional classical counterpart. 
Studies of these QPTs and the quantum-classical mapping rely heavily on the exactly solvable models. 
One of the most common examples is the one-dimensional quantum Transverse Ising Model (1D TIM) \cite{free energy}, which exhibits a second-order QPT at the critical point $\lambda_c=1$, and its 
classical counterpart - the two-dimensional classical Ising model \cite{suzuki}, which exhibits a second-order thermal phase transitions at the Curie point. 

Historically, scaling has played a central role in the study of classical criticality. It is well known that the 2D classical Ising model obeys finite-size scaling behavior \cite {fishier}. A straightforward idea is to study $T\neq 0$ scaling laws of 1D TIM.
In Refs. \cite{thermal} and \cite{heavyfermion}, the authors use Gruneisen Parameter and concurrence to characterize finite-temperature properties induced by QPT at zero temperature. In this paper, instead we will use a classical macroscopic thermodynamic
 obsevable - the MS - to study the finite-temperature properties of the generalized 1D TIM - the quantum XY chain. The MS has the advantage of being easily experimentally accessible and has been used as a witness of macroscopic quantum entanglement \cite{explain,verdral}. We will show how the finite-temperature scaling is manifested when the temperature approaches zero, in analogy with finite-size scaling in the imaginary time direction of the 2D classical Ising model. We will also verify the universality in the properties of the MS in quantum XY chain. Finally we will elucidate the close relation between the MS
and another well studied observable - the GP \cite{pachos,zhu,yi,hamma}.
  
\emph{Magnetic susceptibility of quantum XY chain at finite temperature:} 
The Hamiltonian of quantum XY chain can be written as \cite{free energy}
\begin{equation}
H(\gamma,\lambda)=J \sum_{i=1}^N\left[ \frac{1+\gamma }{2}\sigma _{i}^{x}\sigma
_{i+1}^{x}+\frac{1-\gamma }{2}\sigma _{i}^{y}\sigma _{i+1}^{y}+\lambda
\sigma _{i}^{z}\right], \label{1}
\end{equation}
where $N$ is the number of spins in the chain; $J$ is the coupling strength (for simplicity we choose $J=1$ hereafter); $\lambda$ is external magnetic field, and $\gamma$ describes the anisotropy of the system;
 $\sigma_i^{\alpha}, \alpha=x, y, z$ are the Pauli matrix on the $i$th
site of the chain. After a standard procedure \cite {free energy}, this Hamiltonian can be diagonalized as
$H(\gamma ,\lambda )=\sum_k 2\Lambda_k (\eta_k^{\dagger}\eta_k-1/2)$, where $\eta_k$ is the Fermionic annihilation operator
of the $k$-th mode quasi particle; $\Lambda_k=\sqrt{(\lambda-\cos k)^2+\gamma^2\sin^2 k} $ are one half of the excitation energy for modes $k=2\pi(i-0.5)/N, i=1, 2,\cdots, N/2$. The partition function of the system can be obtained as
$Z=\prod_k\left(e^{-\beta \Lambda_k}+e^{\beta\Lambda_k}\right)=\prod_k 2\cosh\left(\beta\Lambda_k\right)$, where $\beta=1/k_BT$ is the inverse temperature and $k_B$ is the Boltzmann constant.
Accordingly, the free energy per spin of the system can be calculated as $F=-k_BT\ln Z/N=-k_BT\sum_k \ln\left[2\cosh\left(\beta\Lambda_k\right)\right]/N$. In the thermodynamic limit, $N\rightarrow\infty$, we use an integral
to replace the sum and obtain the exact expression of the free energy per spin at temperature $T$ \cite {free energy}
\begin{equation}
F=-k_BT\ln 2-k_BT\times \frac{1}{\pi}\int_{0}^{\pi}dk \ln\left[\cosh\left(\beta \Lambda_k\right)\right]. \label{2}
\end{equation}
The magnetization per spin along the direction of the external magnetic field $\lambda$ at temperature $T$ can be obtained
\begin{equation}
M_z(T)=-\frac{\partial F}{\partial \lambda}=\frac{1}{\pi}\int_0^{\pi}\tanh \left(\beta\Lambda_k \right)\frac{\lambda-\cos k}{\Lambda_k} dk, \label{3}
\end{equation}
and then the MS along z direction $\chi_{z}=-\partial^2F/ \partial \lambda^2$  as a function of the temperature $T$ and the magnetic field $\lambda$ of the system can also be obtained
\begin{equation} 
\chi_{z} (\lambda, T)=\frac{1}{\pi}\int_0^{\pi} \left[   \frac{\beta}{ \cosh^{2}{(\beta \Lambda_{k})}} \frac{(\lambda -\cos k)^{2}}{\Lambda_{k}^{2}} +\tanh(\beta \Lambda_{k})\frac{\gamma^{2} \sin^{2} k}{\Lambda_{k}^{3}}  \right]  dk
\end{equation}

We plot the MS $\chi_{z}$ of 1D TIM ($\gamma=1$) as a function of external magnetic field $\lambda$ and the temperature $T$ in Fig. 1. Clearly it can be seen that the logarithmic divergence
of the MS at zero temperature indicates the second-order QPT at the QCP $\lambda_c=1$. 
We would like to point it out that at zero temperature, the magnetization is reduced to $M_z(T=0)=\int_0^{\pi}(\lambda-\cos k)/(\pi \Lambda_k) dk$. 

For the convenience of later study, we introduce another observable -- the GP, which is a fundamental concept in quantum mechanics \cite{berry}. To obtain a geometric phase, we rotate the Hamiltonian (\ref{1}) around the z axis for an angle $\phi$. The effective Hamiltonian
after the rotation is
\begin{equation}
H_{\phi}=U_{\phi} H U^{\dagger}_{\phi}, \\\\\\\\\\\\\  U_{\phi} =\prod_{j=1}^{N} e^{i \phi \sigma_{j}^{z}/2}. \label{3.1}
\end{equation}
The periodicity of the Hamiltonian in $\phi$ is $\pi$. After we rotate the Hamiltonian back to its initial form ($\phi=\pi$), the GP of the ground state accumulated by varying the angle $\phi$ from 0 to $\pi$ is given by
\begin{equation}
\beta_{g}=-i \frac{2}{N} \int_{0}^{\pi} \left ( \left\langle GS \right\vert U^{\dagger}_{\phi} \right) \frac{\partial}{\partial \phi} \left ( U_{\phi} \left\vert GS \right\rangle \right) d \phi, \label{3.2}
\end{equation}
which is an extra phase in addition to the usual dynamic phase. From Refs \cite{pachos,zhu,yi,hamma} we know that the ground-state GP studied there
can be expressed as
\begin{equation}
\beta_g=\pi+\int_0^{\pi}\frac{\lambda-\cos k}{\Lambda_k} dk = \pi+ \pi M_{z}(T=0). \label{4}
\end{equation}
Hence, the derivative $\partial \beta_g/\partial \lambda$ of the ground-state GP over the external field is $\pi$ times of the zero-temperature MS $\chi_{z}=\partial M_z(T=0)/\partial \lambda$. We can understand this relation
in the following way: the ground-state GP studied in Refs. \cite{pachos,zhu,yi,hamma} is a function of the derivative of the ground state energy with respect to the external magnetic field \cite{zhu,yi,hamma}, and at zero temperature, the free energy is equal to
 the ground state energy. Thus, at zero temperature, the GP is a function of the magnetization.
 
 As is well known, at zero temperature, the MS of 1D TIM shows logarithmic singularity at the QCP and exhibits finite-size scaling behavior in the 
 proximity of the QPT point $\lambda_c=1$. Thus, it is not surprising that the GP exhibits singularity and finite-size scaling behavior near the QCP \cite {zhu}. Instead of studying the finite-size scaling of the GP (MS at 
 zero temperature), in this letter, we will study finite-temperature scaling of the quantum XY chain. We will see that when the temperature approaches zero, in analogy with the imaginary time direction approaching the infinity in the finite-size scaling,
  the MS obeys $T\neq 0$ scaling behavior in the proximity of the QPT. 
\begin{figure}[ht]
\begin{center}
\includegraphics[width=7cm, clip]{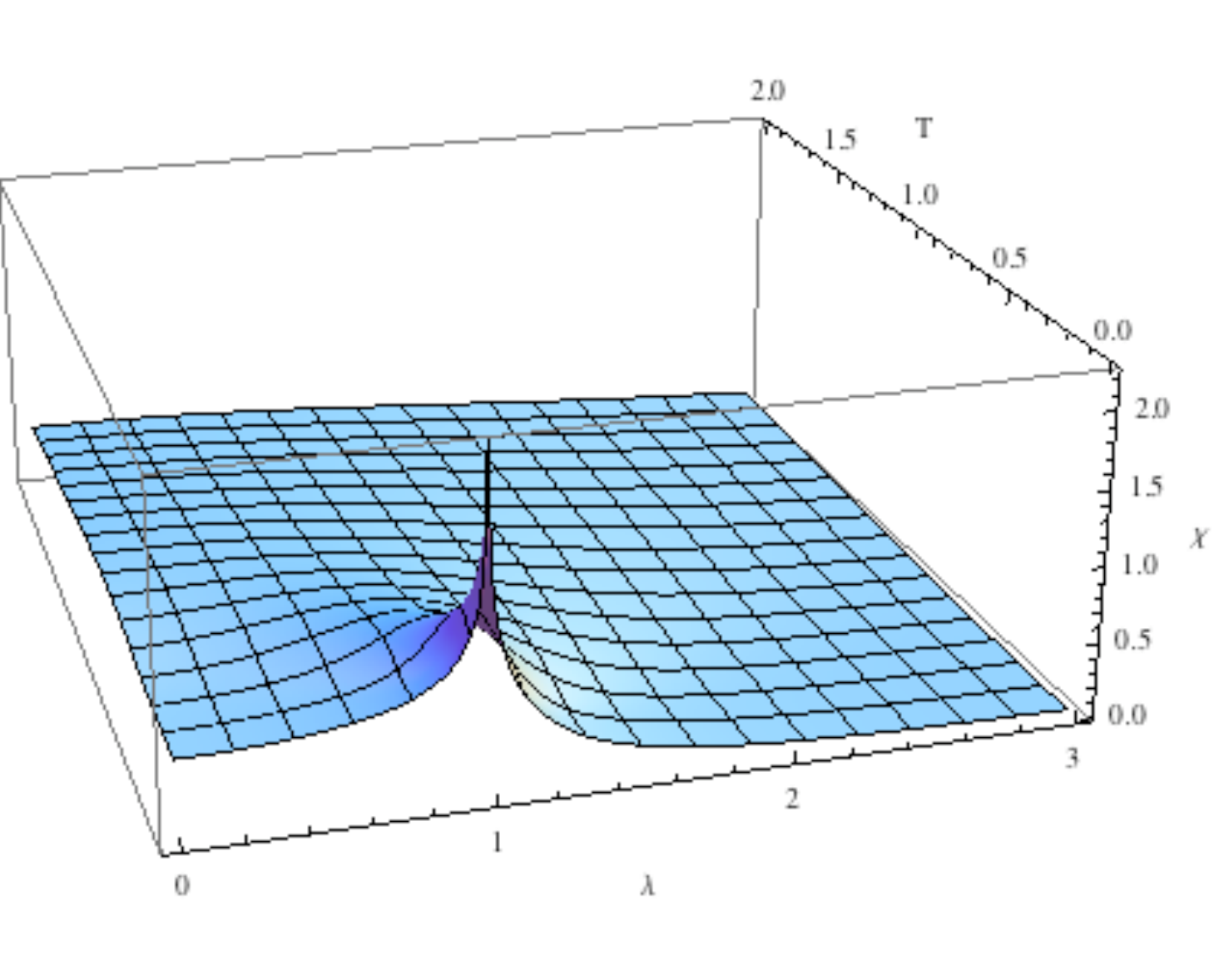}
\end{center}
\caption{MS $\chi_{z}$ of 1D TIM model as a function of external magnetic field $\lambda$ and temperature $T$. It can be seen that the MS at zero temperature
show logarithmic divergence at the QCP $\lambda_c=1$. At nonzero temperature the MS is analytical. This agrees with the known result
that 1D TIM model does not exhibits thermal phase transition at nonzero temerature.}
\label{fig1}
\end{figure}
\begin{figure}[ht]
\begin{center}
\includegraphics[width=6cm, clip]{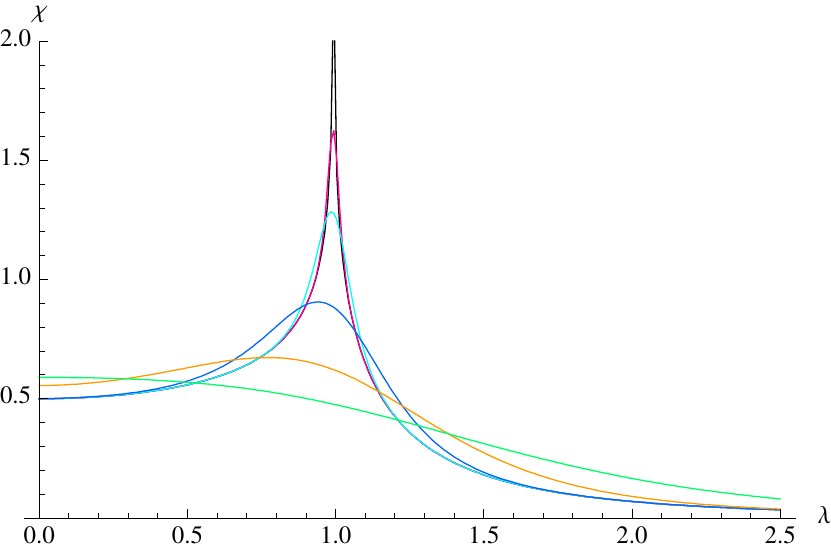}
\includegraphics[width=6cm, clip]{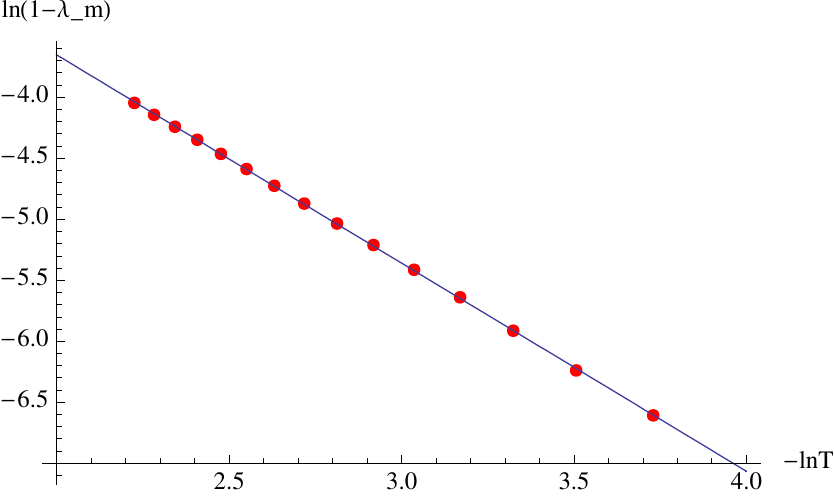}
\end{center}
\caption{(color online). (above) The MS for the 1D TIM ($\gamma=1$) as a function of the controlling parameter $\lambda$. The curve corresponds to different 
temperatures $k_B T=0 \mathrm{J}$, $0.02 \mathrm{J}$, $0.06 \mathrm{J}$, $0.21 \mathrm{J}$, $0.5 \mathrm{J}$, and $1.01 \mathrm{J}$. With the decrease of the temperature,
the maximum gets pronounced, and (below) the pseudopoint $\lambda_{m}$ changes and tends as $T^{1.706}$ towards the QCP $\lambda_c=1$.}
\label{fig1}
\end{figure}

\emph{Scaling of the magnetic susceptibility of the quantum XY chain:} 
In order to further understand the relation between the 1D TIM and 2D classical Ising model, we investigate the finite-temperature scaling behavior of the MS by the finite-size scaling ansatz \cite{barber83}.
For simplicity, we first look at 1D TIM ($\gamma=1$), and we will discuss the properties of the family of $\gamma\neq1$ later. The MSs as a function of the external magnetic field $\lambda$ at different temperatures
$T$ (including zero temperature) are presented in Fig. 2. At zero temperature the MS shows a singularity at $\lambda_{c}=1$, but at nonzero temperature, there are no real divergence of $\chi_{z}$. Nevertheless, there are clear anomalies at low 
temperature, and the height of which increases with the decrease of the temperature. This can be regarded as the precursors of the QPT. What is more, the position $\lambda_m$ of the maximum susceptibility (pseudocritical point) \cite{barber83} changes and tends as
 $T^{1.704}$ towards the QCP and clearly approaches $\lambda_c$ when $T\rightarrow 0$ (see Fig. 2b). Meanwhile, the maximum value $\chi_{z} |_{\lambda_m}$ of the MS diverges logarithmically with the decrease of the temeprature
\begin{equation}
\chi_{z}|_{\lambda_m}\approx\kappa_1\ln{T}+\mathrm{const}.\label{5}
\end{equation}
Our numerical results (see Fig. 3a) give $\kappa_1=0.320$. On the other hand, when $T=0$, from Ref. \cite{barber81} we know that the MS in the proximity of the QCP exhibits logarithmic singularity
\begin{equation}
\chi_{z}\approx\kappa_2\ln{|\lambda-\lambda_c|}+\mathrm{const}.\label{6}
\end{equation}
Our numericals in Fig. 3b give the result $\kappa_2\approx0.317$, while the exact result \cite{barber81} gives $\kappa_2=1/\pi\approx0.3183$.  We would like to point it out that the coefficient $\kappa_2$ here
is the same as that in Ref. \cite{zhu}, where the author gives $\kappa_2\approx0.3123$ and our numerical result is closer to the exact result $\kappa_2=1/\pi$.
\begin{figure}[ht]
\begin{center}
\includegraphics[width=6cm, clip]{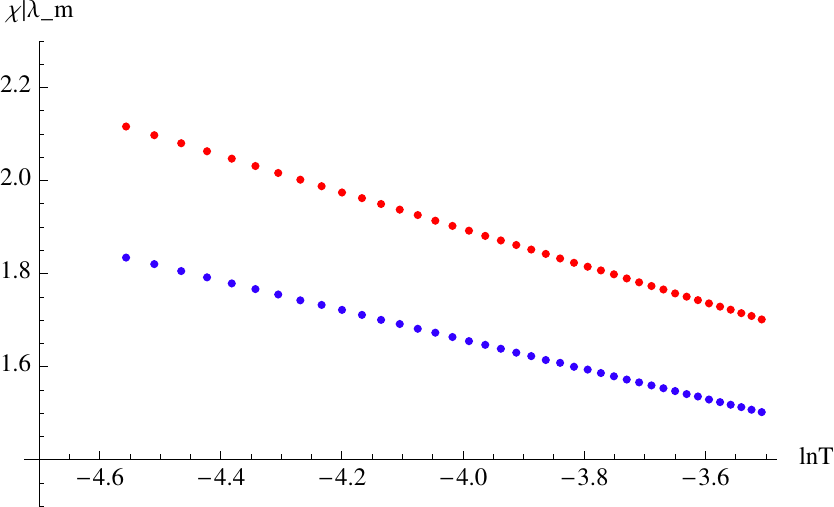}
\includegraphics[width=6cm, clip]{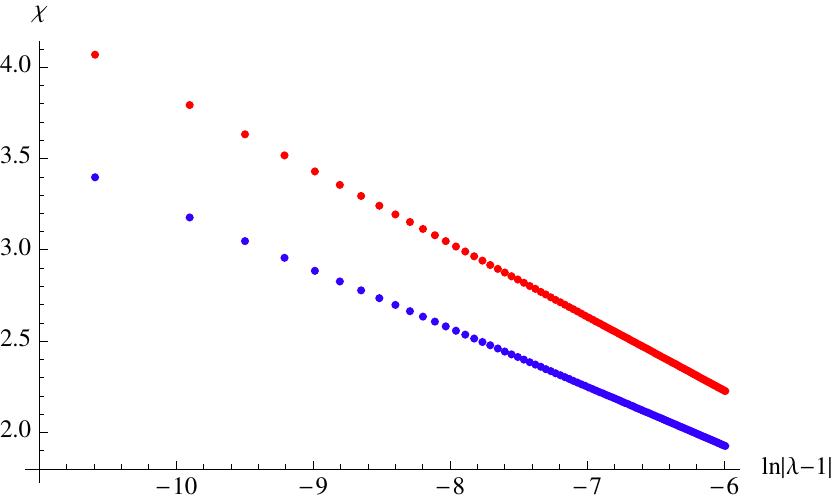}
\end{center}
\caption{(color online). (above) The maximum value of the MS at the pseudocritical point $\lambda_m$ of the 1D quantum XY chain as a 
function of temperature $T$. The slope of the line is $0.317$ ($0.394$) for $\gamma=1$ (blue) ($\gamma=0.8$ (red)). (below) The MS at zero temperature
diverges logarithmically in the proximity of the QCP $\lambda_c=1$. The slope of the line is $0.320$ ($0.401$) for $\gamma=1$ (blue) ($\gamma=0.8$ (red)). The ratio of the two slopes (below and
above) for a fixed parameter $\gamma$ is equal to the critical exponent $\nu$. Here $\nu\approx 1.009$ ($\nu\approx 1.017$) for $\gamma=1$ ($\gamma=0.8$) is obtained. The numerical results
agree with the scaling ansatz and the universality of the XY model.} 
\label{fig3}
\end{figure} 
According to the scaling ansatz in the logarithmic singularities, the ratio $\vert\kappa_2/\kappa_1\vert$ gives the critical exponent $\nu$ that governs the divergence of the correlation length $\xi\sim \vert \lambda-\lambda_c\vert ^{-\nu}$. In our case,
$\nu\approx1.009\sim1$ is obtained in the numerical calculation for the 1D TIM, which agrees well with the known result about 1D TIM \cite{free energy}. Furthermore, by proper scaling and taking into account
the distance of the maximum $\chi_{z}$ from the QCP, it is possible to make all the data for the value of $F=1-\exp\left[\chi_{z}(\lambda)-\chi_{z}\vert_{\lambda_{m}}\right]$ as a function of $(\lambda-\lambda_m)/T$ for different
temperatures $T$ to collapse onto a single curve (see Fig. 4). This figure contains the data for temperatures ranging from $k_BT=e^{-3} \mathrm{J}$, $e^{-4} \mathrm{J}$, $e^{-5} \mathrm{J}$, $e^{-5.5} \mathrm{J}$. These results demonstrate that the MS
does obey the scaling behavior as the temperature decrease to zero, in analogy to the lattice size approaching the infinity in the finite-size scaling cases. 

In the following we will study the universality of the critical behavior of the MS. 
It is well known that the anisotropic XY chain ($\gamma\ni(0, 1]$) belongs to the 1D TIM universality, while isotropic XY chain ($\gamma=0$)
belongs to the XX universality. For the 1D TIM universality, $\nu=1$, while for the XX universality, $\nu=1/2$. We will show
that the finite-temperature scaling behavior of $\chi_{z}$ also manifests the universality principle - the critical properties depends only on the dimensionality of the system and the broken symmetry in ordered phase. 
To verify the universality principle of the XY model, we consider the case for $\gamma\neq1$. The asymptotic behavior is also described by 
Eqs. (\ref{5}) and (\ref{6}). From Fig. 3 we see that for $\gamma=0.8$ numerical simulation gives $\kappa_1\approx0.394$ and $\kappa_2\approx0.401$, while the exact result \cite{barber81} should be $\kappa_2=(\gamma \pi)^{-1}\approx 0.398$.
 As a result the critical exponent for $\gamma=0.8$ is $\nu=\vert\kappa_2/\kappa_1\vert\approx1.017$, very close to the exact value $\nu=1$. Moreover we also verify that by proper scaling, all data for different temperatures $T$ but a specific $\gamma$ will collapse onto the same curve. The data for $\gamma=0.8$ are shown in Fig. 4. 
\begin{figure}[ht]
\begin{center}
\includegraphics[width=8cm, clip]{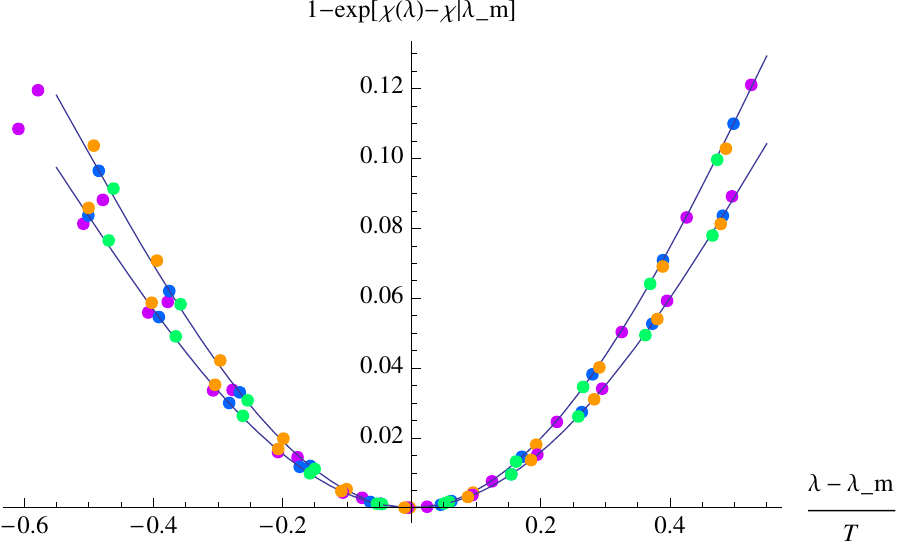}
\end{center}
\caption{(color online). The value of $F=1-\exp{[\chi_{z}(\lambda)-\chi_{z}|\lambda_m]}$ as a function of $(\lambda-\lambda_m)/T$ for
different temperatures (dots of different colors) $k_B T=e^{-3} \mathrm{J}, e^{-4} \mathrm{J}, e^{-5} \mathrm{J}$, and $e^{-5.5} \mathrm{J}$. For fixed $\gamma$ (here we choose
$\gamma=1$ and $\gamma=0.8$), all data collapse on a single curve, which agrees with the finite-size scaling behavior. The critical exponent $\nu=1$ can be 
obtained from this figure.}
\label{fig4}
\end{figure}
What is more, through a similar analysis to that in Ref. \cite{zhu}, we can directly extract the finite-temperature scaling behavior of the XX ($\gamma=0$) universality class. It can be found that, at zero temperature $T=0$, for the XX universality, the magnetization
can be written in the following compact form.
\begin{equation}
M_{z}= \left \{ 
\begin{array}{c} 1 - \frac{2}{\pi} \arccos \lambda , (0\leq \lambda\leq 1)  \\ 
1, (\lambda >1)
\end{array}
\right. .\label{9}
\end{equation}
Accordingly, the critical exponent $\nu=1/2$ and $z=2$ can be extracted from the MS $\chi_{z}=\sqrt {2} (1- \lambda)^{-\frac{1}{2}}$, $(\lambda \to 1^{-})$ \cite{zhu}, which is different from the TIM universality ($\nu=1$ and $z=1$). When we change the anisotropy $\gamma$ from 1 to 0, we find the range of the validity of the quantum scaling ansatz in $\lambda$ (Eq. (\ref {6})) shrink gradually. The leading term of the MS crossover from
$\frac{1}{\pi \gamma} \ln {(1-\lambda)}$ to $\sqrt {2} (1- \lambda)^{-\frac{1}{2}}$. Hence, when $0<\gamma \leq 1$, the scaling belongs to XY universality, while when $\gamma=0$, the scaling belongs to XX universality. Finally, we also would like to point it out that our numerical result shows that the scaling behavior of Eq. (\ref{5}) can persist up to a temperature $k_B T\approx0.5 \mathrm{J}$. This result 
agrees well with that of Ref. \cite {nature05}. In addition, the crossover line in the region $0<\lambda<1$ given by the MS $\chi_{z}$ is roughly $T_c \sim \left\vert \lambda-\lambda_c \right\vert^{\nu z}$, which agrees well with the result obtained in the analysis
elsewhere \cite{thermal}. Hence the boundary of quantum critical scaling region can be confirmed by the behavior of the MS $\chi_{z}$.

\emph{Magnetic susceptibility and geometric phase:} 
As we have mentioned before, the derivative of the ground-state GP discussed in Refs. \cite{pachos,zhu,yi,hamma} is equal to $\pi$ times of the MS, and the finite-size scaling of the GP \cite{zhu} actually represents the finite-size
scaling of the MS. Based on these studies, we would like to further study the relation between the thermal-state GP and the MS at a finite temperature. Similar to the definition
of the ground-state GP in Refs. \cite{pachos,zhu,yi,hamma}, we define the thermal-state GP in the following way: four eigenstates of the modes $(k, -k)$ of $H_{\phi}$ (see Refs. \cite{pachos,zhu,yi,hamma}) can be expressed as $\left\vert 00 \right\rangle_k
=\cos{(\theta_k/2)}\left\vert 0\right\rangle_k\left\vert 0\right\rangle_{-k}+i e^{i2\phi}\sin{(\theta_k/2)}\left\vert 1\right\rangle_k\left\vert 1\right\rangle_{-k}$, $\left\vert 11 \right\rangle_k=
i e^{-i2\phi}\sin{(\theta_k/2)}\left\vert 0\right\rangle_k\left\vert 0\right\rangle_{-k}+\cos{(\theta_k/2)}\left\vert 1\right\rangle_k\left\vert 1\right\rangle_{-k}$, $\left\vert 01 \right\rangle_k=\left\vert 0\right\rangle_k\left\vert 1\right\rangle_{-k}$
, and $\left\vert 10 \right\rangle_k=\left\vert 1\right\rangle_k\left\vert 0\right\rangle_{-k}$ with the angle $\theta_k$ defined by $\theta_k=\arctan [-\sin k/(\cos k-\lambda)]$. The GP of the thermal state at temperature $T$ accumulated by varying 
the angle $\phi$ from $0$ to $\pi$ is described by 
\begin{equation}
\beta_T=\frac{-2i}{N}\sum_{k=1}^{N/2}\sum_{n} \int e^{-\beta E_n^k} \left\langle n\right\vert_k \frac{\partial}{\partial \phi}\left\vert n\right\rangle_k d \phi, \label{7}
\end{equation}
where $\left\vert n\right\rangle_k=\left\vert 00\right\rangle_k, \left\vert 01\right\rangle_k, \left\vert 10\right\rangle_k$, and $\left\vert 11\right\rangle_k$. 
After a straightforward calculation, we obtain the same relation between the magnetization and the GP as that of zero temperature $\beta_T=\pi+ \int_0^{\pi}(\lambda-\cos k)/\Lambda_k \tanh {(\beta \Lambda_k)} dk=\pi[1+M_z(T)]$ (\ref {3}).
Thus we prove that at both zero temperature and nonzero temperature, the GP of the quantum XY chain is a linear function of the magnetization and the derivative of the GP is proportional to the MS. 
The discussions of the finite-temperature scaling of the MS in this letter can be alternatively regarded as the finite-temperature scaling of the GP in the proximity of the QPT point. Finally the close relation
between the GP and the MS does not confined to 1D quantum XY chain. In Ref. \cite{plaslina} the ground-state GP of the Dicke model and its relation to quantum criticality are studied. We would like to point
it out that, similar to the discussions about the 1D XY chain, the ground-state GP of the Dicke model is a linear function of the ground state magnetization $\beta_{g}=\pi(1+\left\langle S_x\right\rangle/N)$, 
where $\left\langle S_x\right\rangle/N$ is the magnetization $M_x$ per spin in Dicke model. Hence the derivative of ground-state GP of the Dicke model is also equal to $\pi$ times of the MS.
Besides the above two examples, it can be proved that for any QPTs driven by an external magnetic field, such as the Lipkin-Meshkov-Glick model \cite{yi} and 1D XXZ model \cite{xxz}, the relation
between the GP and the magnetization still holds true. The proof is give as follows. For those QPTs driven by an external magnetic field, we apply a $\pi$-rotation along the z axis for every spin $U_{\phi} =\prod_{j=1}^{N} e^{i \phi \sigma_{j}^{z}/2}$ to obtain the GP. 
The GP of the ground state can be expressed as (\ref{3.2})
\begin{equation}
\beta_{g}=-i \frac{2}{N}  \int_{0}^{\pi} \left ( \left\langle GS \right\vert U^{\dagger}_{\phi} \right) \left(i \sum_{j}\frac{\sigma_{j}^{z}}{2}\right) \left ( U_{\phi} \left\vert GS \right\rangle \right) d \phi
= \frac{2\pi}{N} \sum_{j}  \left\langle GS \right\vert \frac{\sigma_{j}^{z}}{2}  \left\vert GS \right\rangle =\pi M_{z}(T=0), \label{10}
\end{equation}
where $\left\vert GS \right\rangle$ is the ground state of the Hamiltonian (\ref{1}) before the rotation, and $U_{\phi} \left\vert GS \right\rangle$ is the instantaneous ground state after the rotation for an angle $\phi$.
We know that $\frac{2}{N} \sum_{j}  \left\langle GS \right\vert \frac{\sigma_{j}^{z}}{2}  \left\vert GS \right\rangle$ is the definition of the GP of the ground state. Thus we prove that the GP obtained by applying a rotation around the z axis to 
each spin, is proportional to the magnetization along z axis (similarly if we rotate along x axis, the GP will be proportional to the magnetization along x axis). We also would like to point out that the GP in Eq. (\ref{10}) differs from that in Eq.
(\ref{4}) by a constant $\pi$. This is because the ground state of $U_{\phi} H U_{\phi}^{\dagger}$ has an uncertainty of the global phase. When we choose a proper global phase, we can eliminate the difference between Eq. (\ref{4}) and Eq. (\ref{10}). When we study the 
scaling of $\frac{d \beta_{g}}{d \lambda}$, the difference does not affect. In addition, we can generalize the above discussions to eigenstates other than the ground state. We find the same proportional factor between the GP and the magnetization
for all eigenstates. Thus the relation between the GP and the magnetization can be straightforwardly generalized to a thermal state at finite temperature. The above XY model is a good example.

In summary, we study the finite-temperature scaling of the MS of the quantum XY chain. All key features of the quantum criticality, such
as scaling, critical exponent, the universality, etc. are presented in the MS of the XY spin chain. Though the nature of the QPT and the $T\neq 0$ scaling is purely quantum mechanical, the classical macroscopic
thermodynamic observable MS, which can be easily accessed experimentally, can be used to witness and characterize the quantum features of the system \cite {explain,verdral}. 
Our studies shed light on the mechanism of bring quantum criticality up to a finite temperature, and opens the possibility of observing the footprint of quantum criticality experimentally.
We also would like to point it out that the results obtained in this paper does not depend on the model and thermodynamic observable used here and can be generalized to other QPT 
models with the only change of MS to a ``controlling parameter-dependent susceptibility".
For example, in a QPT driven by the pressure instead of the external magnetic field, the observable $\chi_p=-\partial^2 F/ \partial p^2$ is expected to exhibit the finite-temperature scaling behavior, and the critical 
exponent can be extracted through a similar analysis. Finally, our study establishes the connection between the MS and the GP at both zero temperature and nonzero temperature in a family of spin systems, where
the QPTs are driven by an external magnetic field.
 
The author thanks F. M. Cucchietti and Rishi Sharma for stimulating discussions and gratefully acknowledges the support of the U.S. Department of Energy through 
the LANL/LDRD Program for this work.


\begin{thebibliography}{99}
\bibitem{explain} the use of thermodynamics observables to witness quantum entanglement is studied in Ref. \cite{verdral}. For XY chain, the separable bound of the MS is given by the sum of three susceptibilities along three orthogonal axes $\chi_{x}+\chi_{y}+\chi_{z}\geqslant N (2k_{B}T)^{-1}$. We know that when $0 \leqslant \gamma \leqslant 1$ and at $T=0$, $\chi_{x}\approx (1-\lambda)^{-7/4}$, $\chi_{y}=0$, and $\chi_{z}\approx (\gamma \pi)^{-1} \ln{|1-\lambda|}$. Hence, except for $\lambda=1$, the MS cannot be explained without entanglement at $T=0$. The above result agrees with the analysis using the specific heat as an entanglement witness \cite{verdral}.
\bibitem{xxz}The Hamiltonian of XXZ model is $H=\sum_{i=1}^N\left[ \sigma _{i}^{x}\sigma
_{i+1}^{x}+\sigma _{i}^{y}\sigma _{i+1}^{y}+\Delta \sigma _{i}^{z}\sigma _{i+1}^{z}+\lambda
\sigma _{i}^{z}\right]$. Here both the anisotropy $\Delta$ and the external magnetic field $\lambda$ can induce QPTs \cite{takahashi}.
 However, only the QPT induced by $\lambda$ can be characterized by the GP induced by a rotation $U_{\phi}=\prod_{j=1}^{N}e^{-i \phi \sigma_{j}^{z}/2 }$.
 Because $\frac{\partial F}{\partial \Delta}$ is not proportional to the magnetization in z direction $M_{z}$, and hence is not proportional to $\left \langle \sigma^{z} \right\rangle$.
\end{thebibliography}
\end{document}